\documentclass[twocolumn,english]{revtex4-1}
\usepackage[T1]{fontenc}
\usepackage[latin9]{inputenc}
\usepackage[a4paper]{geometry}
\makeatletter

\providecommand{\tabularnewline}{\\}

\renewcommand\[{\begin{equation}}
\renewcommand\]{\end{equation}}

\makeatother

\usepackage{babel}
\begin{document}

\title{The Effect of Electrical Boundary Conditions on the Thermal Properties
of Ferroelectric Piezoelectric Ceramics}

\author{Husain N.~Shekhani$^{1*}$, Erkan A.~Gurdal$^{2}$, Lalitha Ganapatibhotla$^{3\dagger}$,
Janna K.~Maranas$^{4}$,\\ Ron Staut$^{5}$, Kenji Uchino$^{1}$}

\address{$^{1}$Department of Electrical Engineering, The Pennsylvania State
University, State College, PA 16801, USA \\ $^{2}$The Center for
Dielectrics and Piezoelectrics, The Pennsylvania State University,
State College, PA 16801, USA \\$^{3}$Performance Materials Division,
The DOW Chemical Company, Houston, TX, USA\\ $^{4}$ Department of
Chemical Engineering, The Pennsylvania State University, State College,
PA 16801, USA\\ $^{5}$APC International, Ltd. Duck Run, Mackeyville,
PA, \emph{17750}, USA\inputencoding{latin1}{\\}\inputencoding{latin9}$^{*}$Author
to whom correspondence should be addressed. Electronic mail:hns116@psu.edu\inputencoding{latin1}{\\}\inputencoding{latin9}$^{\dagger}$Author
was previously at the Department of Chemical Engineering at The Pennsylvania
State University }
\begin{abstract}
The thermal conductivity of polycrystalline bulk PZT (lead-zirconate-titanate)
has been investigated according to electrical boundary conditions
and poling. The thermal conductivity of poled PZT was measured in
the poling direction for open circuit and short circuit conditions.
The short circuit thermal conductivity had the largest thermal conductivity.
The relationship between these two thermal properties, the ``electrothermal''
coupling factor $k_{33}^{\kappa}$, was found to be similar to the
electromechanical coupling factor $k_{33}$ relating elastic compliance
under short circuit and open circuit conditions. The thermal conductivity
of the unpoled sample was found to have the lowest thermal conductivity.
The significance of the thermal conductivity with regards to phonon
mode scattering and elastic compliance was discussed.
\end{abstract}
\maketitle
Piezoelectric materials are a unique class of materials in which the
electrical and mechanical properties are coupled. The most popular
piezoelectric material is lead-zirconate-titanate (PZT) due to its
large electromechanical properties and the adaptability of its properties
with dopants.  

It is well known that the electrical boundary conditions and poling
affect the elastic compliance of piezoelectric materials.$^{1}$ The
thermal properties, namely thermal conductivity, may also be expected
to be affected by electrical boundary conditions and poling due to
the fact that this property and elastic compliance both arise primarily
from phonon mode phenomena. 

The thermal properties of PZT ceramics have been studied at low temperature
(20K-300K), and a transition temperature was found between 50K and
80K.$^{2}$ It has also been studied at high temperature, between
300K-800K, which characterizes the affect of phase transition on the
thermal properties.$^{3}$ However, the relationship of thermal properties
with electrical boundary conditions and poling in PZT and other ferroelectric
piezoelectric ceramics has not been studied.

Using the experiment outlined in$^{4}$, which uses a time constant
formulation to determine thermal properties, the thermal diffusivity
$\alpha$ of a poled and depoled commercially available hard PZT ceramic
discs, APC 841 (APC Int., USA), of a diameter of 51 mm was measured.
The thermal diffusivity was measured in the direction of polarization
for the poled samples. Using the DSC Q2000 (TA Instruments), the absolute
value heat capacity of a small sample was measured by comparing it
to a sapphire reference sample. The heat capacity and density are
respectively $c_{p}=340\mathrm{J/kg\,K}$ $\rho=$7600$\mathrm{kg/m^{3}}$.
The heat capacity is not affected by electrical boundary conditions
and it is a scalar property; therefore, the thermal conductivity $\kappa$
can be determined from the thermal diffusivity $\alpha$ and from
the heat capacity $c_{p}$

\begin{equation}
\kappa=\alpha c_{p}\rho.
\end{equation}
The original report regarding the TC thermal diffusivity experiment
presented a standard deviation of less than 5\%. However, the results
of this experiment had larger deviation (Tab.~1) due to the fact
that the diameters of the samples were a few millimeters smaller than
the sample holder. The error found is different for different boundary
conditions and poling states, but this is believed to be random.

\begin{table*}[!t]
\caption{Thermal diffusivity and thermal conductivity depending on electrical
boundary conditions\label{tab:Thermal-diffusivity-and}}

\centering{}%
\begin{tabular}{ccccc}
 &  &  &  & \tabularnewline
\hline 
\hline 
 & Thermal diff. & $+/-$ & Thermal cond. & $+/-$\tabularnewline
 & $\mathrm{(mm^{2}/s)}$ &  & $\mathrm{(W/m\,K)}$ & \tabularnewline
\hline 
Open circuit & 0.50 & 0.02 & 1.4 & 0.06\tabularnewline
Short circuit & 0.82 & .08 & 2.3 & 0.23\tabularnewline
Depoled & 0.43 & 0.03 & 1.2 & 0.01\tabularnewline
\hline 
\hline 
 &  &  &  & \tabularnewline
\end{tabular}
\end{table*}

The measured thermal diffusivity and thermal conductivity values for
poled open circuit, poled short circuit, and depoled samples are described
in Tab.~\ref{tab:Thermal-diffusivity-and} as an average of two measurements
on three samples each. samples were depoled by heating them and then
checking their response on a $d_{33}$ meter. The short circuit thermal
conductivity $\kappa_{33}^{E}$ is more than 1.5 times larger than
the open circuit one $\kappa_{33}^{D}$. The unpoled thermal conductivity
$\kappa^{u}$ showed the smallest value, 15\% less than that of the
open circuit case. The relationship between the open circuit $\kappa_{33}^{D}$
and short circuit thermal conductivity $\kappa_{33}^{E}$ can be described
by a electrothermal coupling coefficient

\begin{equation}
\kappa_{33}^{E}(1-(k_{33}^{\kappa})^{2})=\kappa_{33}^{D},\label{eq:kappa}
\end{equation}
which is closely related to the relationship$^{1}$ between short
circuit and open circuit elastic compliance and the electromechanical
coupling coefficient $k_{33}$ 

\begin{equation}
s_{33}^{E}(1-k_{33}^{2})=s_{33}^{D}.
\end{equation}
The electromechanical coupling factor of this ceramic found from electrical
impedance spectroscopy is $k_{33}=$0.68. Using Eq.~\ref{eq:kappa},
the ``electrothermal'' coupling factor can be calculated to be $k_{33}^{\kappa}=$0.63.
The error between the two coupling factors may be due to error in
the thermal measurements and possibly other microscopic features which
do not correlate between the $k$ value determined from electrical
and thermal measurements.

In summary, $\kappa_{33}^{E}>\kappa_{33}^{D}>\kappa^{u}$. This may
be understood from phonon mode scattering, orientation of domains,
and elastic compliance. Because of the random orientation of domains
in the depoled sample, it is expected that there will be the most
phonon scattering in this material. Therefore, it will have the lowest
thermal conductivity. Because the domains of the poled material are
oriented, less scattering is expected and thermal conductivity will
be larger for the poled material ($\kappa_{33}^{E}$ and $\kappa_{33}^{D}$).
The elastic compliance under short circuit conditions $s_{33}^{E}$
is softer than the elastic compliance under open circuit conditions
$s_{33}^{D}$. This means that the lattice and domain wall motion
are larger in the short circuit condition. The larger lattice vibration
and domain motion in short circuit conditions will also correlate
to a larger thermal conductivity in short circuit conditions $\kappa_{33}^{E}$
due to increased phonon mode transport. This observation can also
be used to understand the relation between $k_{33}$ and $k_{33}^{\kappa}$

The clear result of the experiments is that thermal conductivity in
ferroelectric ceramics depends on electrical boundary conditions.
It is very possible that the electromechanical coupling factor in
these materials is related to thermal properties as well, namely thermal
conductivity and thermal diffusivity. A discussion was presented to
explain the behavior using phonon scattering and domain orientation
concepts. Future work includes studying the effect of microstructure
on phonon mode transport in these materials and further clarifying
the thermal-electrical coupling experimentally demonstrated in the
experiments.

\begin{acknowledgments}
The authors would like to acknowledge the Office of Naval Research
for sponsoring this research under grant number: ONR N00014-12-1-1044.
\end{acknowledgments}

\section*{Bibliography}

\noindent $^{1}$B.~Jaffe, W.R.~Cook Jr., and H.~Jaffe. Piezoelectric
Ceramics, London, (Academic Press, 1971)

\noindent $^{2}$S.~Yarlagadda, M. Chan, H.~Lee, G.A.~Lesieutre,
D.W.~Jensen, and R.~S.~Messer. ``Low temperature thermal conductivity,
heat capacity, and heat generation of PZT.'' \emph{Journal of intelligent
material systems and structures}, 6.6, 757-764 (1995)

\noindent $^{3}$S.N.~Kallaev, G.G.~Gadzhiev, I.K.~Kamilov, Z.M.~Omarov,
S.A.~Sadykov, and L.A.~Reznichenko. ``Thermal properties of PZT-based
ferroelectric ceramics.'' \emph{Physics of the Solid State}, 48.6,
1169-1170 (2006)

\noindent $^{4}$Shekhani, H.N., and Uchino, K. ``Thermal diffusivity
measurements using insulating and isothermal boundary conditions.''
\emph{Review of Scientific Instruments,} 85.1, 015117 (2014)

\end{document}